\documentclass[aps,preprint,preprintnumbers,nofootinbib,showpacs]{revtex4}

\def\beq{\begin{equation}}
\def\eeq{\end{equation}}
\def\beqa{\begin{eqnarray}}
\def\eeqa{\end{eqnarray}}

\def\za{\alpha}
\def\zb{\beta}
\def\ssc{\scriptscriptstyle}
\def\lsim{\mathrel{\raise.3ex\hbox{$<$\kern-.75em\lower1ex\hbox{$\sim$}}} }
\def\gsim{\mathrel{\raise.3ex\hbox{$>$\kern-.75em\lower1ex\hbox{$\sim$}}} }

\begin{document}
\draft
\preprint{\tighten{\vbox{\hbox{NCU-HEP-k015}\hbox{KIAS-P04011}
\hbox{Jan 2004}\hbox{rev. Jun 2004}
}}}
\vspace*{1in}

\title{Little Higgs Model Completed with a Chiral Fermionic Sector
\vspace*{.3in} 
}
\author{\bf Otto C. W. Kong 
\vspace*{.2in}
}
\email{otto@phy.ncu.edu.tw}
\affiliation{ Department of Physics, National Central University, Chung-li, Taiwan 32054
\\ \& \ \ \ \ Korea Institute for Advanced Study, Seoul, Korea
}

\begin{abstract}
The implementation of the little Higgs mechanism to solve the hierarchy problem provides 
an interesting guiding principle to build particle physics models beyond the electroweak
scale. Most model building works, however, pay not much attention to the fermionic sector.
Through a case example, we illustrate how a complete and consistent fermionic sector 
of the TeV effective field theory may actually be largely dictated by  the gauge structure
of the model. The completed fermionic sector has specific flavor physics structure,
and many phenomenological constraints on the model can thus be obtained beyond
gauge, Higgs, and top physics. We take a first look on some of the quark sector 
constraints.
\end{abstract}
\pacs{}
\maketitle


The hierarchy problem, or the very un-natural fine-tuning required to fix  the electroweak scale 
due to the quadratic divergent quantum corrections to the Higgs boson mass, is a major
theoretical short-coming of the Standard Model (SM). The fine-tuning problem can be alleviated, 
only if there is new physics at the TeV scale that guarantees the cancellation of the quadratic 
divergence to an acceptable level, or totally changes our picture of SM physics. A guaranteed 
cancellation has to come from some mechanism protected by a symmetry. Candidates of the 
kind include supersymmetry, and the recently proposed little Higgs mechanism
\cite{dd,S}. 

With the little Higgs idea, the SM Higgs boson is to be identified as pseudo-Nambu Goldstone boson(s) 
(PNGB) of some global symmetries. Two separate global symmetries, each to be broken by a Higgs 
vacuum expectation value (VEV), are to be arranged such that a 1-loop (SM) Higgs mass diagram is 
protected by the (residue) symmetries to be free from quadratic divergence. The idea was motivated by 
dimensional deconstruction\cite{dd}, though the mechanism may not necessarily follows from the strong 
interaction dynamics behind (see Ref.\cite{L}). Simple group theoretical constructions of little 
Higgs models have also been proposed\cite{S,KS}. In this article, we take the perspective of
considering a little Higgs model as an effective field theory at the TeV scale and look into some 
plausible implications on quark physics.

A little Higgs model typically has an extended electroweak gauge symmetry, with extra fermions,
at least a top-like quark $T$, carrying nontrivial gauge charges. Such extra fermions have to be
vectorlike at the SM level. Chiral states cannot have mass much above the electroweak scale 
and are hence extremely dangerous phenomenologically. Individual chiral fermions also ruin
the gauge anomaly cancellation within the SM; and adding more fermions to restored the
consistency is far from a trivial business. The extra fermions vectorlike under the SM
symmetry are, however,  likely to be of chiral nature before the breaking
of the extra gauge symmetry. The  $T$ quark has to be connected to the SM $t$ quark by the 
symmetries that enforce the quadratic divergence cancellation. For any such symmetry to be
compatible with the SM symmetries, $t_{\!\ssc L}$ and  $t_{\!\ssc R}$ cannot belong to multiplets
of the same gauge charged; hence, it would likely be the same for $T_{\!\ssc L}$ and  $T_{\!\ssc R}$. 
Having fermions of fundamentally chiral nature also naturally fix their mass at or below the 
corresponding gauge symmetry breaking scale. From the lesson of the SM itself, one can see that the 
chiral fermionic spectrum embodies the beauty of the model, and dictates many of the properties of the 
fermions. In fact, it has been argued that the spectrum can be derived from the gauge anomaly
cancellation requirement\cite{unc67}. The latter is the best one can do answering the 
question of why there is what there is in a particle physics model up to now. We present here
such a chiral fermionic spectrum as a consistent completion of a little Higgs model and take
a first look into the resulted implication on quark physics. We are working in a specific
model, however, similar issues should be relevant to all little Higgs models though the exact
fermionic completion and other details would be model dependent.

Our background little Higgs model is a model with  $SU(3)_C\times SU(3)_L\times U(1)_X$
gauge symmetry given in Ref.\cite{KS}. The model has problem with the quartic Higgs coupling,
which can be fixed in a $SU(3)_C\times SU(4)_L\times U(1)_X$ extension\cite{KS}. Here,
we mainly stick to the $SU(3)_L$ version for simple illustration. Generalizing to the $SU(4)_L$
version is mostly straight forward. Moreover, an alternative little Higgs model with the same
gauge symmetry is given in Ref.\cite{ST}, to which most of the results here will likely apply. Our
focus here is to illustrate the basic features a little Higgs model with a complete and consistent 
fermionic content could have. In our opinion, the perspective could take little Higgs model 
construction and phenomenological studies to a new level. We skip discussion on the scalar and 
gauge boson sector, as well as the working of the little Higgs mechanism itself, for which readers are 
referred to the original reference\cite{KS}. Within the model, the $T_{\!\ssc L}$ forms a $SU(3)_L$ 
triplet with the SM $(t, b)$ doublet, while $T_{\!\ssc R}$ is a singlet, same as  $t_{\!\ssc R}$. 
The first point to note here is that one cannot embed the other two families of SM quarks in 
the same way. Unlike $SU(2)_L$, $SU(3)_L$ multiplets are not free from triangle anomaly.
In fact, the existence of a phenomenologically viable anomaly free embedding for the full content 
of the SM fermions together with $T_{\!\ssc L}$ and  $T_{\!\ssc R}$  is not {\it a priori} obvious.
Solving the problem more or less dictates the properties of the admissible complete model.
Interestingly enough, a simple solution exists\cite{090}. We present the spectrum in Table I
and a similar result for the $SU(4)_L$ model in Table II. We emphasize again the results
are not arbitrary choices of anomaly free spectra, they are essentially the minimal chiral
spectra satisfying the requirement. Moreover, in the case of the one-family SM spectrum, the 
anomaly cancellation conditions tied the multiplets together so closely that no (simple) 
alternative is possible. It is expected to be the same case here. It is also of interest to note
that the embedding of the three families of SM fermions are different, with gauge anomaly
cancellation among them. In this sense, the consistent spectrum has to contain three families,
which can be taken as a way to understand why there is a triplication of the anomaly free
spectrum for the SM itself. 

\begin{table}[b]
\noindent
\caption{\small
 Fermion spectrum for the $SU(3)_C\times SU(3)_L\times U(1)_X$ model with 
little Higgs. Here, we give the hypercharges of the electroweak states, with SM doublets 
put in [.]'s. The other states are singlets.
The embedding of the electric charge is given by 
${\cal Q}= \frac{1}{2} \, \lambda^3 
 - \frac{1}{2\sqrt{3}} \, \lambda^8 + X$ (with $\rm{Tr}\{\lambda^a\lambda^b\}=2\delta^{ab}$).
}
\begin{center}
\begin{tabular}{|c|cc|}
\hline\hline
				& \multicolumn{2}{|c|}{$U(1)_Y$-states}			\\  \hline
${\bf (3_{\scriptscriptstyle C},3_{\scriptscriptstyle L} ,\frac{1}{3})}$                 &  ${\bf \frac{1}{6}}$[$Q$]               & ${\bf \frac{2}{3}}$($T$)       	\\
2\ ${\bf ({3}_{\scriptscriptstyle C},\bar{3}_{\scriptscriptstyle L} ,0)}$     &  2\ ${\bf \frac{1}{6}}$[2\ $Q$]               & 2\ ${\bf \frac{-1}{3}}$($D,S$)     \\
$3\ {\bf (l_{\scriptscriptstyle C} ,3_{\scriptscriptstyle L} ,\frac{-1}{3})}$           & 3\  ${\bf \frac{-1}{2}}$[3\ $L$]          	& 3\ {\bf 0}(3\ $N$)   \\
$4\ {\bf (\bar{3}_{\scriptscriptstyle C},1_{\scriptscriptstyle L} ,\frac{-2}{3})}$        & \multicolumn{2}{|c|}{4\ ${\bf \frac{-2}{3}}$ ($\bar{u}, \bar{c}, \bar{t}, \bar{T}$)} 	     	\\
$5\ {\bf (\bar{3}_{\scriptscriptstyle C},1_{\scriptscriptstyle L} ,\frac{1}{3})}$        & \multicolumn{2}{|c|}{5\ ${\bf \frac{1}{3}}$ ($\bar{d}, \bar{s}, \bar{b}, \bar{D}, \bar{S}$)} 	                 	\\
$3\ {\bf (1_{\scriptscriptstyle C},1_{\scriptscriptstyle L} ,1)}$              &    \multicolumn{2}{|c|}{3\ ${\bf 1}$  ($e^+, \mu^+, \tau^+$)   }      	\\
\hline\hline
\end{tabular}
\end{center}
\end{table}

\begin{table}[b]
\noindent
\caption{\small 
Fermion spectrum for a direct $SU(3)_C\times SU(4)_L\times U(1)_X$ extension
of the model (see Ref.\cite{KS} for the little Higgs structure). 
Again, we give the hypercharges of the electroweak states, with 
SM doublets put in [.]'s. The embedding of the electric charge is given by 
${\cal Q}= \frac{1}{2} \, \lambda^3 
 - \frac{1}{2\sqrt{3}} \, \lambda^8 +  - \frac{1}{2\sqrt{6}} \, \lambda^{15} +X$
}
\begin{center}
\begin{tabular}{|c|ccc|}
\hline\hline
				& \multicolumn{3}{|c|}{$U(1)_Y$-states}			\\  \hline
${\bf (3_{\scriptscriptstyle C},4_{\scriptscriptstyle L} ,\frac{5}{12})}$                 &  ${\bf \frac{1}{6}}$[$Q$]               & ${\bf \frac{2}{3}}$($T$)  &      ${\bf \frac{2}{3}}$($T'$) 	\\
2\ ${\bf ({3}_{\scriptscriptstyle C},\bar{4}_{\scriptscriptstyle L} ,\frac{-1}{12})}$     &  2\ ${\bf \frac{1}{6}}$[2\ $Q$]               & 2\ ${\bf \frac{-1}{3}}$($D,S$)     & 2\ ${\bf \frac{-1}{3}}$($D',S'$) \\
$3\ {\bf (l_{\scriptscriptstyle C} ,4_{\scriptscriptstyle L} ,\frac{-1}{4})}$           & 3\  ${\bf \frac{-1}{2}}$[3\ $L$]          	& 3\ {\bf 0}(3\ $N$) 		 & 3\ {\bf 0}(3\ $N'$)   \\
$5\ {\bf (\bar{3}_{\scriptscriptstyle C},1_{\scriptscriptstyle L} ,\frac{-2}{3})}$        & \multicolumn{2}{|c}{4\ ${\bf \frac{-2}{3}}$ ($\bar{u}, \bar{c}, \bar{t}, \bar{T}$)} 	& { ${\bf \frac{-2}{3}}$ ($\bar{T'}$)} 	          	\\
$7\ {\bf (\bar{3}_{\scriptscriptstyle C},1_{\scriptscriptstyle L} ,\frac{1}{3})}$        & \multicolumn{2}{|c}{5\ ${\bf \frac{1}{3}}$ ($\bar{d}, \bar{s}, \bar{b}, \bar{D}, \bar{S}$)}     & 2\ ${\bf \frac{1}{3}}$ ($\bar{D'}, \bar{S'}$)                  	\\
$3\ {\bf (1_{\scriptscriptstyle C},1_{\scriptscriptstyle L} ,1)}$              &    \multicolumn{2}{|c}{3\ ${\bf 1}$  ($e^+, \mu^+, \tau^+$)   }      	 &   	\\
\hline\hline
\end{tabular}
\end{center}
\end{table}

With the spectrum, we look into the possible couplings of the two $SU(3)_L$ Higgs multiplets  
$\Phi_i$, as given in Ref.\cite{KS},  to the fermions. One should note that the content and quantum
numbers of Higgs multiplets are a central, non-negotiable, feature of the little Higgs model.
The couplings are responsible for the SM Yukawa couplings, and the basic properties of the
extra singlet quarks, as well as the leptons. We give below details of the quark sector Yukawa 
couplings, assuming that the lowest order terms admitted by the gauge symmetry are all allowed. 
The discussion is to illustrate explicitly that the model at least do admit sensible Yukawa
couplings for the SM quarks, with the extra, SM singlet, quarks generally getting masses
from the $SU(3)_L\times U(1)_X \to SU(2)_L\times U(1)_Y$ symmetry breaking. We
will comment on the resultant mass matrices result in view of experimental data below.

First comes the top sector. It is given in Ref.\cite{KS} as
\beqa 
{\mathcal L}_{top} \!\! &=& \!\!\!\!  \lambda^t_{\scriptscriptstyle 1}\,\bar{t}^\prime_a\,
\Phi_{\!\scriptscriptstyle 1} \, Q^a   +    \lambda^t_{\scriptscriptstyle 2}\,\bar{T}^\prime_a\,
\Phi_{\!\scriptscriptstyle 2} \, Q^a
\nonumber \\  =\!\!\!\! && \!\!\!\!
f\,(\lambda^t_{\scriptscriptstyle 1}\,\bar{t}^\prime+  
\lambda^t_{\scriptscriptstyle 2}\,\bar{T}^\prime)\, T
+ \frac{i}{\sqrt{2}} \, (\lambda^t_{\scriptscriptstyle 1}\,\bar{t}'
-  \lambda^t_{\scriptscriptstyle 2}\,\bar{T}^\prime)\,h
\left( \begin{array}{c} t \\ b \end{array} \right) + \cdots
\nonumber \\ =\!\!\!\! && \!\!\!\!
m_{\scriptscriptstyle T}\, \bar{T} T - i y_{\!\scriptscriptstyle t} \, \bar{t}\, h
\left( \begin{array}{c} t \\ b \end{array} \right) + \cdots\;. \label{Lt}
\eeqa
Here $Q^a$ denotes the triplet of $t$, $b$, and $T$ quarks (these are chiral fermionic states;
here and below, we suppress the $L$ or $R$ subscripts);
and we suppress the color indices ($a$) after the first line. Both $\lambda^t_{\scriptscriptstyle 1}$ 
and $\lambda^t_{\scriptscriptstyle 2}$ are expected to be of order one to produced the 
phenomenological top mass from electroweak symmetry breaking. In fact, we have 
$m_{\scriptscriptstyle T}=\sqrt{(\lambda^t_{\scriptscriptstyle 1})^2+ 
(\lambda^t_{\scriptscriptstyle 2})^2} \, f$ and $y_{\!\scriptscriptstyle t}
=\sqrt{2} \lambda^t_{\scriptscriptstyle 1} \lambda^t_{\scriptscriptstyle 2} /
 \sqrt{(\lambda^t_{\scriptscriptstyle 1})^2+ (\lambda^t_{\scriptscriptstyle 2})^2}\;$.
A piece not explicitly given in the above top-sector (or rather up-sector) Yukawa couplings
 to the triplet $Q$ is the term
\beq \label{Tt}
\frac{-i}{\sqrt{2}} \, 
 \frac{(\lambda^t_{\scriptscriptstyle 2})^2- (\lambda^t_{\scriptscriptstyle 1})^2}
{(\lambda^t_{\scriptscriptstyle 1})^2+ (\lambda^t_{\scriptscriptstyle 2})^2} \;
\bar{T} \,  h \left( \begin{array}{c} t \\ b \end{array} \right) \;.
\eeq
This term represents deviation of up-sector
quark physics from that of the SM, hence, deserves attention\cite{Sf}. A nonzero value of the 
term, in fact, also signifies deviations of  $m_{\scriptscriptstyle T}$ and 
$y_{\!\scriptscriptstyle t}$ given above from what the notation suggests. There are actually 
more mixings of the SM quarks with the electroweak singlet $T$, as shown below.

We have two more SM quark doublets residing in the  $\bar{3}_{\!\scriptscriptstyle L}$ 
representations $Q^\prime_j$ ($j=1$ and $2$). There are admissible dimension five terms
\beqa 
{\mathcal L}_{Q^\prime} &=&  \frac{1}{M} \lambda^u_{\za j} \, \bar{u}^\prime_\za \,
\Phi_{\!\scriptscriptstyle 1} \, \Phi_{\!\scriptscriptstyle 2} \,  Q^\prime_j  
\\ \nonumber  	&=&
\frac{-i \, \sqrt{2} \, f}{M} \lambda^u_{\za j} \, \bar{u}^\prime_\za \, h \, 
\left( \begin{array}{c} u_j \\ d_j \end{array} \right) + \cdots \;,
\label{LQ'} \eeqa
where color indices are suppressed (same below), with $M$ being a background mass scale 
factor and $\lambda^u_{\za j} $  a $4\times 2$ matrix of couplings with obvious indexing 
for the quark states.  Note that in the generic case, all four quark singlets have to be allowed to
couple to $\Phi_{\!\scriptscriptstyle 1} \, \Phi_{\!\scriptscriptstyle 2} \,  Q^\prime_j $.
However, we are still left with an $SU(2)$ flavor degree of freedom of basis choice
among the singlets besides $\bar{t}$ and $\bar{T}$ and an another $SU(2)$ flavor
basis choice among the two $Q^\prime_j$'s. Hence, we can give an optimal parametrization
of the up-sector mass matrix as
\beq
{\cal M}^u = \left( \begin{array}{cccc}
m_{\!\scriptscriptstyle u}^{\prime}	& 0 & 0 & 0 \\
0 & m_{\!\scriptscriptstyle c}^{\prime}	 & 0 & 0 \\
m_{\!\scriptscriptstyle t\!u} & m_{\!\scriptscriptstyle t\!c} & m_{\!\scriptscriptstyle t}^{\prime} 	& 0\\
m_{\!\scriptscriptstyle T\!u} & m_{\!\scriptscriptstyle T\!c} &
m_{\!\scriptscriptstyle T\!t} & m_{\scriptscriptstyle T} 
\end{array} \right)   \;.
\eeq
The $m_{\!\scriptscriptstyle Tt}$ is from the coupling as given by expression (\ref{Tt}),
which we discussed. The other mass mixing terms are all from the dimension five terms
involving the  $Q^\prime_j$'s. 

Next, we look at the down-quark sector. The bottom quark has to get its Yukawa coupling from 
the dimension five $\bar{b}\,\Phi_i^\dag \Phi_j^\dag \, Q$ term, which may naturally give 
the desired suppression in its mass. For the first two families, however, the 
$1_{\!\scriptscriptstyle L} \, \Phi_i^\dag \, \bar{3}_{\!\scriptscriptstyle L} $ terms
are naively allowed for all the down type quarks. Putting all that together, we have the 
following dimension four and five terms responsible for the quark masses :
\begin{widetext}
\beqa
{\mathcal L}_{down}  &=&  \lambda^{d1}_{\zb j} \, \bar{d}^\prime_\zb \,
\Phi_{\!\scriptscriptstyle 1}^\dag \,   Q^\prime_j  + \lambda^{d2}_{\zb j} \, \bar{d}^\prime_\zb \,
\Phi_{\!\scriptscriptstyle 2}^\dag \,  Q^\prime_j   + \frac{1}{M} \lambda^b_{\zb} \, \bar{d}^\prime_\zb \,
\Phi_{\!\scriptscriptstyle 1} ^\dag\, \Phi_{\!\scriptscriptstyle 2} ^\dag\,  Q
\\ \nonumber  	&=& 
f\,( \lambda^{d1}_{\zb j} \, \bar{d}^\prime_\zb +   \lambda^{d2}_{\zb j} \, \bar{d}^\prime_\zb) \, D_j
- \frac{i}{\sqrt{2}} \, (\lambda^{d1}_{\zb j} \, \bar{d}^\prime_\zb
 -   \lambda^{d2}_{\zb j} \, \bar{d}^\prime_\zb)\,
h^\dag  \left( \begin{array}{c} u_j \\ d_j \end{array} \right) 
+  \frac{i \, \sqrt{2} \, f}{M} \lambda^b_{\zb} \, \bar{d}^\prime_\zb  \, h^\dag \, 
\left( \begin{array}{c} t \\ b\end{array} \right) + \cdots 
 \label{Ld} \eeqa
\end{widetext}
We note that the $\zb$ index goes from 1 to 5. The basic notation should be obvious.
The first term can be used to extract the two states within the $SU(5)$ flavor space of the
five quark singlets as $\bar{D}_j$'s, the singlets that couple directly to the $D_j$'s.
However, one cannot then avoid having the couplings of the $\bar{D}_j$'s in the latter
two terms simultaneous without extra assumption. The mass matrix for the down-sector 
quarks  may then be written in the $3+2$ block form
\beq
{\cal M}^d = \left( \begin{array}{cc}
m^{\!\scriptscriptstyle d} & 	0 \\
m^{\!\scriptscriptstyle D\!d} & m^{\!\scriptscriptstyle D}
\end{array} \right) 	\;,
\eeq
where we are leaving $m^{\!\scriptscriptstyle d}$ and $m^{\!\scriptscriptstyle D\!d}$
as generic matrices to stick to the left-handed basis of the SM doublets as in ${\cal M}^u$ 
above and to accommodate the required nontrivial CKM mixings of the SM quarks.

It is clear from the above that the quantum number assignment scheme does admit
heavy, $SU(3)_L\times U(1)_X \to SU(2)_L\times U(1)_Y$ symmetry breaking scale,
masses to the extra singlet quarks, together with Yukawa couplings for all the SM quarks. 
There are also mass mixings between the two classes, as parametrized above by
$m^{\!\scriptscriptstyle T\!u} = \left( m_{\!\scriptscriptstyle T\!u} \;\; 
m_{\!\scriptscriptstyle T\!c} \;\; m_{\!\scriptscriptstyle T\!t} \right)$ and 
$m^{\!\scriptscriptstyle D\!d}$. At least in the limit that these mixings are vanishingly
small, one will obtain SM quark physics with the heavy quarks decoupled. This result
is, of course, what is to be expected naively. We have only to emphasize that there does
exist a large enough number of independent couplings, in the Lagrangian parts, as given
by Eqs.(\ref{Lt},\ref{LQ'},\ref{Ld}) to make the scenario admissible. It should also be
noted that the couplings as given here do not automatically  produce the full
hierarchical quark mass pattern. 

The physics of the type of extra vectorlike quarks have been studied under various
scenarios before \cite{vlq}. We will take a step in the direction to get an idea on some of the
constraints on the model. This is mainly an attempt to illustrate the strong phenomenological
implications the fermion sector of a little Higgs model may have, as well as to outline the 
particular type of phenomenological constraints to be expected for the current model
under discussion. We leave a detailed phenomenological study along the line to 
future publications.

The first set of stringent constraint on a SM extension with vectolike quarks mixing with 
the SM ones is the precision results on partial widths of the $Z$-decay. The heavy quarks 
have to be above the decay threshold; but the modified electroweak nature of the SM 
quarks changes the partial widths of the latter. The effective coupling of an 
electroweak state $f$ to the $Z^{\ssc 0}$ boson is proportional to 
$T^3_{\!f} - {\cal Q}_{\!f} \, \sin\!^2\theta_{\!\ssc W}$
from which one can easily worked out the mass eigenstate couplings. In our case, 
the extra quarks are all electroweak singlets. The first order mixings are among
the $L$-handed states. Since these are mixings among states of different electroweak
character, non-universal flavor diagonal couplings as well as flavor changing neutral
current (FCNC) couplings are induced. We introduce diagonalizing matrices for 
${\cal M}^{u\dag}{\cal M}^{u}$ and  ${\cal M}^{d\dag}{\cal M}^d$ in $3+1$ and $3+2$ 
block form  as
\beq
U^f_{\!\ssc L} =
\left( \begin{array}{cc}
K^f	&	R^f	\\
S^f	&	T^f
\end{array} \right)	\;,
\eeq
for $f=u$ and $d$. Obviously, admissible mixings between the SM $L$-handed quarks and
the heavy $L$-handed singlet quarks have to be very small. A block perturbative analysis 
then  yields the solution :
\beqa
R^d &\simeq& m^{{\!\ssc D\!d} {\dag}} \, (m^{{\!\ssc D }{\dag}})^{-1} \; T^d  \;,
\nonumber \\
S^d &\simeq& - (m^{\!\ssc D})^{-1} \;  m^{\!\ssc D\!d}  \, K^d  \;,
\eeqa
and
\beqa
R^u &\simeq& \frac{1}{m_{\ssc T}} \;  m^{{\!\ssc T\!u} {\dag}} 	\;,
\nonumber \\
S^u &\simeq& \frac{-1}{m_{\ssc T}} \;  m^{\!\ssc T\!u}  \, K^u  \;.
\eeqa
Here, the $K^f$ and $T^f$ matrices are essentially the unitary matrices that diagonalize the 
corresponding diagonal blocks ($T^u$ is just the unit element 1). 

Couplings of $L$-handed SM quark mass eigenstates to the $Z^{\ssc 0}$ boson is
modified to
\beqa
g_{\!\ssc L} (u) &=& \frac{1}{2}  \left[ 1- |S_u|^2  \right] - \frac{2}{3}  \, \sin\!^2\theta_{\!\ssc W} \;,
\nonumber \\
g_{\!\ssc L} (c) &=& \frac{1}{2} \left[ 1- |S_c|^2  \right] - \frac{2}{3}  \, \sin\!^2\theta_{\!\ssc W}  \;,
\eeqa
 for the $u$ and $c$ quark, where the $(S_u \;\;S_c\;\; S_t)$ denote the $1\times 3$ matrix $S^u$,
and for the down-sector 
\beqa
g_{\!\ssc L} (q) &=& - \frac{1}{2}  \left[ 1- |(S^d)_{1q}|^2 - |(S^d)_{2q}|^2 \right] 
+ \frac{1}{3}  \, \sin\!^2\theta_{\!\ssc W} 
\nonumber \\
&& \qquad\qquad (\rm{for \ } q= d,\;s,\;\rm{and\ }b)\;\;.
\eeqa
Similarly, the induced FCNC couplings are given by expression of the form
\beq
g_{\!\ssc L} (\bar{u}c) = \frac{1}{2}  \left[ - S_u^*S_c  \right] \;,
\eeq
for example.

Applying the above results against the experimental data, we can get an idea on the 
constraints on the very nontrivial flavor structure of the model. In particular all the 
$g_{\!\ssc L}(q)$ results decrease as a result of the $S^u$ and $S^d$ mixings. Current data 
\cite{Lang}  allows roughly only a decrease of the total hadronic width by $0.115\%$.
The magnitude of even a single dominant element of the $S^u$ and $S^d$ matrices would
then be bounded roughly by $0.014$, which reflexes the allowable order of magnitude
for a mass ratio of the form $\frac{\rm mixing mass}{\rm heavy mass}$, such as 
$\frac{m_{\!\ssc Tc}}{m_{\ssc T}}$. The up-sector is particularly interesting, as the kind
of mixings most probably exist for any little Higgs model. From Eq.(\ref{LQ'}), we can see that
the required suppression here is actually not too bad, for a mass term like 
${m_{\!\ssc Tc}}$ would naturally be at scale $\sim \frac{f}{M} m_t$ or below 
(if the coupling is suppressed). Mixings $<0.014$ also imply that FCNC couplings
induced are largely safe. For example, $g_{\!\ssc L} (\bar{u}c)$ contributions to $D$-meson 
mixing requires only
\beq
|S_u^* S_c| 
\lsim  2\,\frac{\cos\!\theta_{\!\ssc W}  \, M_{\!\ssc Z}}{g_2} 
\frac{1}{f_{\!\ssc D}} 
\left(\frac{3\,\Delta m_{\!\ssc D}}{2\, m_{\!\ssc D} \, B_{\!\ssc D}} \right)^{\!\! 1/2}
\simeq 0.001 \;,
\eeq
which is not stronger.

A little Higgs model other than the example case here may not contain extra vectorlike
down-type quarks. So, similar constraints from the down-sector are less generic. Otherwise,
the constraints are no weaker. For example, we have $|(S^{d\dag} S^d)_{ds}|$  roughly
bounded by $3\times 10^{-4}$ from kaon physics, without taking into consideration CP phase 
dependent bounds. On the whole, the $<0.014$ bound on light heavy quark mixings is the 
major guideline to be taken.

Readers will realize that our discussion of the phenomenological constraints above  is
quite generic. They are performed on parameters within the quark mass mixing matrices 
rather than explicit model parameters. Of course the mass mixing parameters come from the
Lagrangian parts illustrated. It should be obvious  that the explicit connection is difficult to 
be addressed analytically though. There is ambiguity in the implementation of CKM mixings
onto the quark mass matrices ${\cal M}^{u}$ and  ${\cal M}^d$, for example. Our goal here
is simply to illustrate the kind of strong phenomenological implications the nontrivial
flavor structure, dictated by the fermionic spectrum, have. A few comments on their role
and relation to the other electroweak constraints are in order.

Short of a detailed numerical study, we can still take a look into the likely implications of 
the constraints above on the various parameters of our model at hand. The $\lambda^{t}_{i}$'s  
have to be order one. Mixing among the physical $t$ and $T$ quark will be 
naturally of order $m_t/m_{\!\scriptscriptstyle T}$. The lighter two quarks of the
sector have their masses from the dimension five term of Eq.(\ref{LQ'}), and hence
suppressed by an $f/M$ factor. Further suppressions from the $\lambda^{u}_{\za j}$ 
couplings are needed to get the right masses for the $u$ and $c$ quarks. Natural 
values for the mixings $S_u$ and $S_c$ are expected
to be $m_u/m_{\!\scriptscriptstyle T}$ and $m_c/m_{\!\scriptscriptstyle T}$. Hence,
they look fine. In summary, the small mixings are ``natural" if the singlet states are heavy 
and the mixing mass terms are of the same order as the light masses. This is the case
for the up-sector. The story for the down-sector, however, is more complicated.

The mixings of the heavy $D$ quarks with the SM quarks are expected to be more 
alarming. This is especially true in the case of the bottom quark.
To appreciate that better, let us recall the basic admissible couplings as given
in Eq.(\ref{Ld}). The bottom quark mass eigenvalue is to come mainly from the 
dimension five term,  $\frac{1}{M} \lambda^b_{\zb} \, \bar{d}^\prime_\zb \,
\Phi_{\!\scriptscriptstyle 1} ^\dag\, \Phi_{\!\scriptscriptstyle 2} ^\dag\,  Q$
while those of the strange and down quarks the dimension four terms
$\lambda^{d1}_{\zb j} \, \bar{d}^\prime_\zb \,
\Phi_{\!\scriptscriptstyle 1} ^\dag \,   Q^\prime_j $  and
$\lambda^{d2}_{\zb j} \, \bar{d}^\prime_\zb \,
\Phi_{\!\scriptscriptstyle 2} ^\dag \,   Q^\prime_j $.  The latter are also the
source of the heavy $D$ quark masses and their mixings with $d$ and $s$. The SM
quark mass hierarchy then dictates, naively, very small values for the couplings 
$\lambda^{d1}_{\zb j}$ and $\lambda^{d2}_{\zb j}$. This is in a big contrast to
the up-sector situation an $f/M$ factor helps to suppressed the lighter quark masses,
with independent Yukawa terms responsible for the heavier quark masses.
Here, the $f/M$ factor suppresses the $b$ mass, relative to the $t$, but enhances
the $d$ and $s$ masses relative to the $b$ itself. We have then a dilemma. We
need small couplings to get the right $d$ and $s$ masses. We need, however,
relative big masses for the extra singlet $D$ states. The latter being light enough
to add extra channels to the hadronic $Z$-width is far too dangerous. For instance,
the effective $s$ quark Yukawa coupling has to be $\sim 10^{-3}$, while an
effective $D$ quark Yukawa of the same order would give a $D$ mass of
 $\sim 10^{-3} f$ --- namely at the GeV order. There is a possible way out though.
Consider phenomenologically admissible effective Yukawa couplings for the $D$
quarks. Couplings of the required magnitude may be restricted to involve the
$\bar{D}$ states among the right-handed singlets. One will have to tune the 
couplings among $\lambda^{d1}_{\bar{D} j}$ and $\lambda^{d2}_{\bar{D} j}$
[{\it cf.} Eq.(\ref{Ld})]  to get small enough values of  mixing masses with $d$
and $s$ --- essentially, the magnitude of 
$(\lambda^{d1}_{\bar{D} j} -\lambda^{d2}_{\bar{D} j}) \, v/f$. The combinations
$(\lambda^{d1}_{\bar{D} j} -\lambda^{d2}_{\bar{D} j})$ are exactly the 
couplings responsible for the $D$ quark masses. Finally, one
has to take simply small $\lambda^{d}$-type couplings for the $\bar{d}'$
states orthogonal to the $\bar{D}$ states to get the right $d$ and $s$ masses.
All in all, we see that the model most probably does have parameter space
regions that can pass the flavor sector constraints discussed here. It does, however,
impose strong requirement on the couplings, especially that of the down-sector.
Relatively light singlet $D$ quarks are preferred. More detailed analysis,
together with numerical studies on the flavor physics of the model should be
performed.

The stringent  fermion sector constraints do not stand along phenomenologically. 
Contributions to FCNCs from quark mixings have to be considered together with the
corresponding contributions from the heavy gauge bosons ($Z'$ and $W'$) 
exchanges. The two type of FCNC contributions typically come into the same
processes. A realistic analysis has to combine the two parts together.
There are also other precision electroweak constraints largely
complementary to the FCNC ones. For instance, Ref.\cite{KS} claims a lower bound
on $f$ of about 1.5 TeV as a result. 

We have a specific model here that contains vectorlike quarks of both the up- and down-type,
which generally mix with the SM quarks. There are very stringent constraints on the admissible
mixings. What we want to emphasize, however, is that any realistic little Higgs model
completed with a consistent fermionic spectrum is likely to contain extra quarks. There 
has to be certainly an extra top-like quark which mixes not only with the top, but most
probably with the up and charm too.  If a consistent chiral spectrum can be found, 
anomaly cancellation requirements would likely dictate the existence of other SM 
singlet fermions. At lease with the example(s) at hand, there are also extra down-sector 
quarks and leptons. This has a strong implication on flavor physics structure from which
interesting constraints can be obtained besides the constraints on the gauge and
Higgs sectors. More detailed analyses of the interplay of all the constraints for a realistic
model should be taken seriously. In particular, it will be interesting to check if the more
realistic $SU(4)_L\times U(1)_X$ model may satisfy the FCNC constraints more
naturally.

We illustrate above, with the case example, how a complete and consistent 
fermionic sector of the TeV effective field theory may actually be largely 
dictated by the gauge structure of the model. While the specific solution
spectrum construction strategy can hardly be generalized to little Higgs
models with an extended electroweak gauge symmetry beyond the 
$SU(N)\times U(1)$ type, the paramount importance of the gauge anomaly
cancellation constraints and the plausible implication of a solution
fermionic spectrum are generic. The latter may be in contrary to the 
impression one may get from the literature on little Higgs, as the author
seems to be the only one drawing attention to the issue so far. Especially because
of that, we want to emphasize and elaborate further on the point here. 

In particular, let us take a look at a $SU(5)/SO(5)$ little Higgs 
model\cite{llh}, which is arguably the most popular one around. In this
case, the electroweak symmetry is to be extended with an extra $SU(2)$, or 
$SU(2)\times U(1)$ factor. The only extra fermionic state explicitly discussed
in the paper is the heavy top quark, with its full gauge quantum
numbers not explicitly stated. Naively, one may be led to the simple choice of a
vectorlike singlet under the extended electroweak symmetry. The choice looks like 
there is no need for any further fermionic states and the model is completed. This is 
actually not what the original authors has in mind, as clearly indicated
by the sentence, we quote, "We do not concern ourselves with the cancellation 
of the $G_1\times G_2$ anomalies in this low energy effective theory, since
there may be additional Fermions at the cutoff which cancel the anomalies
involving the broken subgroup."\cite{llh} So, the gauge anomaly issue
is recognized, but pushed aside instead of solved.
 
In our opinion, the simple vectorlike singlet choice is not really quite
feasible, nor desirable. Moreover, the idea of pushing any additional 
fermions to the cutoff scale may not be in much better shape either.

The authors of Ref.\cite{llh} did have the anomalies issue in mind. Only fermions 
chiral with respect to the full gauge symmetry contribute to the anomalies.
Mass for a chiral fermion is by definition ruled out by gauge invariance, 
while a vectorlike pair has admissible Dirac mass naturally at the cutoff scale. 
If the heavy top is vectorlike, before any gauge symmetry breaking, its mass would
likely be at the cutoff $M$. It is the chiral states which only match into vectorlike pair 
of the broken gauge symmetry that should be expected to have mass at the scale
$f$ or below. Chiral fermions, rather than vectorlike ones, are what is more
relevant to low energy physics. It is possible to make the extra fermions heavy by
adoption nonperturbatively large Yukawa couplings, provided that they do form 
vectorlike pairs with respect to the SM gauge group. It is however not very appealing
to have mass terms forbidden by some gauge symmetry to be larger than the gauge
invariant mass terms.

The quantum numbers of a full gauge multiplet containing any SM
multiplet or the heavy top states are dictated by the symmetry embedding 
of gauge group into the parent (global) $SU(5)$. While the version of the model with
only an extra $SU(2)$ gauge symmetry is formally consistent having only the
vectorlike $T$ singlet, so long as one does not give up the idea of a possible $SU(5)$ 
description of the fermion sector, the existence of extra fermionic states charged under
the extra gauge symmetry is not a matter of arbitrary choice. It is
a model consistency issue to be looked into carefully. For a generic model,
the existence of such a consistent fermionic spectrum is not {\it a priori}
guaranteed. It is not just about adding states to cancel the anomalies. One
has also to make sure that the resulted spectrum, when split into SM multiplets,
gives no other chiral electroweak states beyond that of the three SM families that
will run into conflict with phenomenology. And it will be of great interest to see if
there are other fermions beyond the heavy top to enrich the prediction of such a 
model at the TeV scale.

A little Higgs model is supposed to describe a TeV scale effective 
theory. A so-called UV completion model of strong dynamics is expected
to be behind the cutoff. Independent of any little Higgs model, it
does not sound likely at all that one with the minimal fermion spectrum
of the SM parts plus only one extra, vectorlike, top quark would arise. We certainly
hope that a more interesting spectrum would be obtained.

After all, the SM fermionic spectrum is fully chiral, and (for a single 
family) essentially dictated by the gauge anomaly cancellation conditions.
This gives an explanation for why the spectrum is what it is, as well
as the light masses of the resulted Dirac fermions. The only state 
within the SM that can have a gauge invariant mass term before 
electroweak symmetry breaking is the Higgs. The latter is then the only 
possible source of the electroweak scale. It is exactly the stabilization
puzzle of this scale that the little Higgs idea aims at resolving.
An all round appealing little Higgs model, in our opinion, should be
one which maintains all these nice features of the SM and, hopefully,  
provides some insight on the origin of the three SM families. We illustrate 
a case example with some partial success in the direction. It should be very 
interesting to see if any other little Higgs model can be similarly completed with 
a chiral fermionic sector. Successful fermionic completion makes a
little Higgs model a more compelling candidate theory beyond the SM.
The kind of flavor physics constraints outlined here above then will
likely play an important role in the experimental checking of the model.

The author thanks the hospitality of Institute of Physics, Academia Sinica, Taiwan,
during the early phase of the work. 
His work is partially supported by the National Science Council of Taiwan, under
grant number NSC92-2112-M-008-044.


\end{document}